\documentstyle[aps,preprint,eqsecnum,floats,epsf]{revtex}
\tighten
\draft
\begin{document}

\preprint{}
\title{Symplectic Symmetry of the Neutrino Mass and the See-Saw
  Mechanism}
\author{A.B. Balantekin\thanks{Electronic address: {\tt
baha@nucth.physics.wisc.edu}}}
\address{Department of Physics, University of Wisconsin\\
         Madison, Wisconsin 53706 USA\thanks{Permanent Address},\\
and\\
Max-Planck-Institut f\"ur Kernphysik,
Postfach 103980, D-69029 Heidelberg, Germany}

\author{N. \"Ozt\"urk\thanks{Electronic address: {\tt
nozturk@science.ankara.edu.tr}}} \address{Department of Physics,
University of Wisconsin\\ Madison, Wisconsin 53706 USA,\\ and\\
Department of Engineering Physics, Faculty of Sciences, Ankara
University,\\ Ankara 06100, Turkey}

\maketitle

\begin{abstract}
We investigate the algebraic structure of the most general neutrino
mass Hamiltonian and place the see-saw mechanism in an algebraic
framework. We show that this Hamiltonian can be written in terms of
the generators of an $Sp(4)$ algebra. The Pauli-G\"{u}rsey
transformation is an $SU(2)$ rotation which is embedded in this
$Sp(4)$ group. This $SU(2)$ also generates the see-saw mechanism. 
\end{abstract}

\pacs{}

\newpage

\vglue1cm

\section{Introduction}

Neutrinos play an essential role in electroweak physics and there is a
fascinating interplay between neutrino properties and various
astrophysical phenomena. Recent observations of especially the
zenith-angle dependence of atmospheric neutrinos \cite{atm}, and the
solar neutrino flux \cite{solar} provide strong hints of non-zero
neutrino masses and oscillations. Atmospheric and solar neutrino
experiments along with direct experimental limits on the neutrino mass
indicate that neutrino masses are much smaller than the
charged-lepton masses. In the Standard Model neutrinos are
left-handed. Hence the only Lorentz scalar pairing is the Majorana
mass which is a weak isotriplet violating global lepton number
conservation. Since the Standard Model has no isotriplet Higgs to
couple to such a neutrino pair neutrinos are taken to be
massless. Thus the neutrino mass is perhaps the most exquisite hint
for physics beyond the Standard Model.  In the current wisdom the
Standard Model is not a fundamental, but an effective theory. Weak
gauge bosons, W and Z, couple at low energies only with left-handed
interactions. Neutrinos, being neutral leptons, interact only weakly
at low energies. Hence the right-handed components of the neutrinos do
not interact, that does not mean that they do not exist. A fundamental
understanding of the origin and the magnitude of neutrino mass and
mixings is lacking, however the see-saw mechanism \cite{seesaw} gives
a natural scheme for the smallness of the neutrino masses and provides
a convenient ansatz for model building. A very nice discussion of
massive neutrino scenarios beyond the standard model is given in
Reference \cite{ramond}.

Approaches to the neutrino mass based on symmetries are not fully
exploited. Algebraic formulation of a problem could provide a check of
the consistency of the formalism and may help to uncover
symmetries. In particular it could be beneficial to rewrite the
neutrino  mass term in the Hamiltonian in terms of the elements of a
Lie algebra which also generates the see-saw mechanism.  The purpose
of the present paper is to explore this possibility to place the
neutrino mass matrix and the see-saw mechanism in the same algebraic
framework. 

One can view the mass term in the Hamiltonian as a generalized pairing
problem. For a single neutrino family (four Dirac components) the most
general pairing algebra is $SO(8)$ (see for example \cite{pere}). We
show that for the neutrino mass Lorentz invariance constraints this
algebra to be $Sp(4)$. We also show that the Pauli-G\"ursey
transformation is an $SU(2)_{\rm PG}$ rotation which is embedded in
the  associated $Sp(4)$ Lie group. The see-saw Hamiltonian is
generated by a particular Pauli-G\"ursey transformation. The mass
Hamiltonian sits in the $Sp(4)/SU(2)_{\rm PG} \times U(1)$ coset where
the $U(1)$ is a chirality transformation. 

In the next section we introduce the $Sp(4)$ algebra for a single
neutrino flavor and write the  mass Hamiltonian in terms of its
generators. In Section 3 we show that the Pauli-G\"ursey
transformation belongs to the associated group and obtain the see-saw
mechanism as a particular Pauli-G\"ursey transformation. We also
discuss how the algebraic structure generalizes for the case of many
flavors of neutrinos. Finally a brief discussion of our results
conclude the paper. 

\vskip 1cm

\section{The Algebraic Structure of the Neutrino Mass Matrix}

We shall use the chiral representation of Dirac matrices, in which
\begin{equation}
\vec{\gamma}=\left(\begin{array}{cc}0&\vec{\sigma} \\ \\
-\vec{\sigma}&0 
\end{array}\right)\; ,\;
\gamma_5=\left(\begin{array}{cc}1&0 \\ \\ 0&-1
\end{array} \right) \; ,\;
\gamma_0= \left(
\begin{array}{cc}0&-1 \\ \\ -1&0 \end{array} \right)
\end{equation}
where the matrices $\vec{\sigma}$ are the usual $2\times2$ Pauli
matrices. Since we want to introduce charges we will work with the
Hamiltonians instead of Hamiltonian densities. Introducing the 
left-and right-handed fields 
\begin{equation}
\psi_L=\frac{1}{2}(1-\gamma_5)\psi \;\; ,\;\;
\psi_R=\frac{1}{2}(1+\gamma_5)\psi
\end{equation}
the Dirac-mass term in the Hamiltonian can be written as 
\begin{equation}
H_m^D= m_D\int d^3{x} (\bar{\psi}_L\psi_R+h.c)
\end{equation}
and the most general Majorana mass term as
\begin{equation}
H_m^M=H_m^L+H_m^R=\frac{1}{2}m_L\int d^3{x}(\bar{\psi}_L\psi_L^c+h.c)
+\frac{1}{2}m_R\int d^3{x} (\bar{\psi}_R\psi_R^c+h.c) .
\end{equation}
In these equations $m_D$, $m_L$, and $m_R$ are the Dirac, left-handed
and right-handed Majorana masses and the charge conjugation of $\psi$
field is defined by 
\begin{equation}
\psi^C= C \bar{\psi}^T
\end{equation}
where $C$ is the charge conjugation matrix. In the chiral
representation it is given by 
\begin{equation}
C =\left(\begin{array}{cc}-i\sigma_2&0 \\ \\ o & i \sigma_2 
\end{array}\right).
\end{equation}

Introducing the charges
\begin{mathletters}
\begin{eqnarray}
D_+ &=&\int d^3{x} (\bar{\psi}_L\psi_R) \\ D_-&=&\int
d^3{x}(\bar{\psi}_R\psi_L) = D_+^\dagger
\end{eqnarray}
\end{mathletters}
the canonical equal-time anti-commutation relations between the
fermion fields lead to 
\begin{equation}
[D_+,D_-]= 2D_0
\end{equation}
where $D_0$ is defined as
\begin{equation}
D_0=\frac{1}{2}\int d^3{x}(\psi_L^{\dag}\psi_L-\psi_R^{\dag}\psi_R). 
\end{equation}
We see that $D_+$, $D_-$  and $D_0$ satisfy $SU(2)$ commutation
relations:
\begin{equation}
[D_+,D_-]=2D_0 \;\;,\;\; [D_0,D_+]=D_+ \;\;,\;\;[D_0,D_-]=-D_-.
\end{equation}

Similarly introducing the left-handed 
\begin{mathletters}
\begin{eqnarray}
L_+&=&\frac{1}{2}\int d^3{x}(\bar{\psi}_L\psi_L^c) \\
L_-&=&\frac{1}{2}\int d^3{x}(\bar{\psi_L^c}\psi_L) = L_+^\dagger \\
L_0&=&\frac{1}{4}\int d^3{x}(\psi_L^{\dag}\psi_L-\psi_L\psi_L^{\dag})
\end{eqnarray}
\end{mathletters}
and the right-handed charges
\begin{mathletters}
\begin{eqnarray}
R_+&=&\frac{1}{2}\int d^3{x} (\bar{\psi_R^c}\psi_R) \\
R_-&=&\frac{1}{2}\int d^3{x} (\bar{\psi}_R\psi_R^c) = R_+^\dagger \\
R_0&=&\frac{1}{4}\int d^3{x}(\psi_R \psi_R^{\dag}-\psi_R^{\dag}\psi_R)
\end{eqnarray}
\end{mathletters}
one can see that they also satisfy $SU(2)$ algebra structure:
\begin{equation}
\label{left}
[L_+,L_-]=2 L_0 \;\;,\;\; [L_0,L_+]=L_+ \;\;,\;\; [L_0,L_-]=-L_{-}\\
\end{equation}
\begin{equation}
\label{right}
[R_+,R_-]=2 R_0 \;\;,\;\; [R_0,R_+]=R_+ \;\;,\;\;[R_0,R_-]=-R_{-}
\end{equation}
Also note that $D_0\equiv L_0+R_0$. 

It is easy to show that the $SU(2)$ algebras generated by the left-
and right-handed charges commute:
\begin{equation}
\left[L_{-,+,0},R_{-,+,0}\right]=0 .
\end{equation}
Calculating all the commutation relations between $D$'s, $L$'s and
$R$'s to find a closed algebra structure provides a new set of
generators:  
\begin{mathletters}
\begin{eqnarray}
\left[D_+,L_0\right]&=&-\frac{1}{2} D_+,\\ \left[D_+,L_+\right]&=&0,\\
\left[D_+,L_-\right]&=& A_+,\\
\left[D_+,R_0\right]&=&-\frac{1}{2}D_+,\\ \left[D_+,R_+\right]&=&0,\\
\left[D_+,R_-\right]&=&A_-,\\ \left[D_-,L_0\right]&=&\frac{1}{2}D_-,\\
\left[D_-,L_+\right]&=&-A_-,\\ \left[D_-,L_-\right]&=&0,\\
\left[D_-,R_0\right]&=&\frac{1}{2}D_-,\\
\left[D_-,R_+\right]&=&-A_+,\\ \left[D_-,R_-\right]&=&0,
\end{eqnarray}
\end{mathletters}
where the additional generators are
\begin{mathletters}
\begin{eqnarray}
A_+ = \int d^3{x} \left[ -\psi_L^T C\gamma_0\psi_R \right],\\ A_-= \int
d^3{x} \left[ \psi_R^{\dag} \gamma_0 C (\psi_L^{\dag})^T\right].
\end{eqnarray}
\end{mathletters}
The commutation relations between them give another SU(2) algebra
\begin{equation}
[A_+,A_-]=2(R_0-L_0) \equiv 2A_0 \;\;,\;\; [A_+,A_0]=-A_+ \;\;,\;\;
[A_-,A_0]=A_-.
\end{equation}
Note that $A_0$ is proportional to the neutrino number operator.  The
commutation relations between $A_+$ and all the other generators
\begin{equation}
\label{2.17}
[A_+,D_+] = 2R_+,  \;\;,\;\   [A_+,D_-]=-2L_-,
\end{equation}
\begin{equation}
\label{2.18}
 [A_+,L_0] = \frac{1}{2}A_+, \;\;,\;\  [A_+,L_+]=D_+  \;\;,\;\
 [A_+,L_-]=0, 
\end{equation}
\begin{equation}
\label{2.19}
 [A_+,R_0] = -\frac{1}{2}A_+,  \;\;,\;\ [A_+,R_+]=0,  \;\;,\;\
 [A_+,R_-]=-D_- 
\end{equation}
along with the Hermitian conjugates of the commutators in
Eqs. (\ref{2.17}), (\ref{2.18}), and (\ref{2.19}) close the algebra:
the ten independent generators; $D_+$, $D_-$, $L_+$, $L_-$, $L_0$,
$R_+$, $R_-$, $R_0$, $A_+$, and $A_-$ form a Lie algebra, the
symplectic algebra $Sp(4)$. 

In terms of these generators the most general mass term of the
Hamiltonian can be written as 
\begin{equation}
\label{2.12}
H_{m}=m_D(D_++D_-)+m_L(L_++L_-)+m_R(R_++R_-).
\end{equation}
In writing Eq.(\ref{2.12}) we ignored the phases of the masses. These
can easily be incorporated as e.g. $m_L(e^{i\chi}L_++L_-e^{-i\chi})$. 

One can write down a number of 4-dimensional matrix realizations of
the Sp(4) algebra. Here we present two of them. The first one can be
expressed using the $2\times2$ Pauli matrices as :
\begin{equation}
D_0=\frac{1}{2}\left(\begin{array}{cc}\sigma_3&0 \\ \\ 0&\sigma_3
\end{array}\right) \; ,\;\;\;
D_+=\frac{1}{2}\left(\begin{array}{cc}0&\sigma_+ \\ \\ \sigma_+&0
\end{array}\right) \; ,\;\;\;
D_-=\frac{1}{2}\left(\begin{array}{cc}0&\sigma_- \\ \\ \sigma_-&0
\end{array}\right), 
\end{equation}
\begin{equation}
\vec{L}=\frac{1}{2}\left(\begin{array}{cc}0&0 \\ \\ 0&\vec{\sigma}
\end{array}\right) \; ,\;\;
\vec{R}=\frac{1}{2}\left(\begin{array}{cc}\vec{\sigma}&0 \\ \\ 0&0
\end{array}\right),
\end{equation}
\begin{equation}
A_+=\frac{1}{2}\left(\begin{array}{cc}0&\sigma_3+1\\ \\\sigma_3-1 &0
\end{array}\right)\; ,\;\;
A_-=\frac{1}{2}\left(\begin{array}{cc}0&\sigma_3-1\\ \\\sigma_3+1 &0
\end{array}\right).
\end{equation}
In these expressions $1$ is the $2\times2$ unit matrix. The second
matrix realization is given as
\begin{equation}
\label{Ds}
D_0=\frac{1}{2}\left(\begin{array}{cc}1&0 \\ \\ 0&-1
\end{array}\right) \; ,\;\;\;
D_+=\left(\begin{array}{cc}0&\sigma_1 \\ \\ 0&0
\end{array}\right) \; ,\;\;\;
D_-=\left(\begin{array}{cc}0&0 \\ \\ \sigma_1&0
\end{array}\right);
\end{equation}
\begin{equation}
\label{Ls}
L_0=\frac{1}{4}\left(\begin{array}{cc}\sigma_3+1&0 \\ \\ 0&-\sigma_3 -
1 \end{array}\right) \; ,\;\;\;
L_+=\frac{1}{2}\left(\begin{array}{cc}0&\sigma_3+1 \\ \\ 0&0
\end{array}\right) \; ,\;\;\;
L_-=\frac{1}{2}\left(\begin{array}{cc}0&0 \\ \\ \sigma_3+1&0
\end{array}\right), 
\end{equation}
\begin{equation}
\label{Rs}
R_0=\frac{1}{4}\left(\begin{array}{cc}-\sigma_3+1&0 \\ \\ 0&\sigma_3 -
1 \end{array}\right) \; ,\;\;\;
R_+=\frac{1}{2}\left(\begin{array}{cc}0&-\sigma_3+1 \\ \\ 0&0
\end{array}\right) \; ,\;\;\;
R_-=\frac{1}{2}\left(\begin{array}{cc}0&0 \\ \\ -\sigma_3+1&0
\end{array}\right), 
\end{equation}
\begin{equation}
\label{As}
A_+=\frac{1}{2}\left(\begin{array}{cc}\sigma_-&0\\ \\ 0& -\sigma_+
\end{array}\right)\; ,\;\;
A_-=\frac{1}{2}\left(\begin{array}{cc}\sigma_+&0\\ \\ 0 & -\sigma_-
 \end{array}\right).
\end{equation}
Since there is only one four-dimensional representation of $Sp(4)$
these two matrix realizations can be transformed into one another by
the unitary transformation 
\begin{equation}
U=\left(\begin{array}{clcr}0&0&0&1 \\ \\ 0&1&0&0 \\ \\ 0&0&1&0 \\ \\
1&0&0&0 \end{array}\right). 
\end{equation}
Using the representation given in Eqs. (\ref{Ds}), (\ref{Ls}),
(\ref{Rs}), and (\ref{As}) one can also write $H_m$ of
Eq. (\ref{2.12})  in matrix form
\begin{equation}
\label{massmatrix}
H_{m}=\left(\begin{array}{clcr}0&0&m_L&m_D \\ \\ 0&0&m_D&m_R \\ \\
m_L&m_D&0&0 \\ \\ m_D&m_R&0&0 \end{array}\right). 
\end{equation}
Most discussions of the neutrino mass matrix start with
Eq. (\ref{massmatrix}) and its generalization to many flavors (see,
for example Ref. \cite{boris}. A careful discussion which pays
particular attention to the phases is given in Ref. \cite{haxste}). 

$Sp(4)$ is isomorphic to $Spin(5)$ which is the universal covering
group of $SO(5)$. Representations of $Spin(5)$ can be considered as
spinor representations of $SO(5)$. It may be interesting to point out
that these spinor representations can be realized using quaternions
or, equivalently, using Pauli matrices.
The four dimensional fundamental representation
of $Sp(4)$ is explicitly given by Eqs. (\ref{Ds}) through (\ref{As}). 
To exhibit the $SU(2)$-doublet structure of this representation we
rewrite these equations as
\begin{equation}
D_{\pm} = \frac{1}{2} \sigma_1 \otimes \sigma_{\pm},
\end{equation}
\begin{equation}
\vec{L} = \frac{1}{4}(1 + \sigma_3 ) \otimes \vec{\sigma},
\end{equation}
\begin{equation}
\vec{R} = \frac{1}{4}(1 - \sigma_3 ) \otimes \vec{\sigma},
\end{equation}
and
\begin{equation}
A_{\pm}= \left[ \sigma_{\mp} \otimes \frac{1}{4}(1 + \sigma_3 )
\right] \oplus \left[ \sigma_{\pm} \otimes \frac{1}{4}(\sigma_3 -1 )
\right] .
\end{equation}
In these equations the notation $A \otimes \sigma_1$, etc. stands for
the $4 \times 4$ block matrix where the elements of $\sigma_1$ are 
replaced by the
$2 \times 2$ matrix $A$. Full details and the representation theory of
spinor representations is covered in standard texts (see for example
\cite{murnaghan}).

\section{The Pauli-G\"{u}rsey transformation and the see-saw
mechanism}

In 1957, W. Pauli introduced \cite{pauli} a 3-parameter group which
relates a Dirac spinor to its charge-conjugation. Later F. G\"{u}rsey
showed \cite{gursey} that this group can be extended to particles with
mass and charge. The Pauli-G\"{u}rsey transformation is given by the
3-parameter transformation
\begin{equation}
\psi\rightarrow\psi^{\prime}=a\psi+b\gamma_5\psi^c
\end{equation}
where the parameters $a$ and $b$ satisfy the unitarity condition 
\begin{equation}
|a|^2+|b|^2=1. 
\end{equation}
In this section we elaborate on the relationship between the
Pauli-G\"{u}rsey transformation and the neutrino mass matrix. 

As the discussion in the previous section indicates the physical
meaning of the $SU(2)$ algebras generated by the $D$'s, $L$'s, and
$R$'s is clear. They generate the neutrino mass matrix. To illustrate
the role of the $SU(2)$ algebra spanned by $A_{\pm}$ and $A_0$ let us
consider a general element of the associated $SU(2)$ group
\begin{equation}
\label{3.3}
\hat{U}=e^{-\tau^* A_-}e^{-log(1+|\tau|^2)A_0}e^{\tau A_+}e^{i\varphi
A_0},
\end{equation}
where $\tau$ and $\varphi$ are arbitrary complex and real numbers
respectively. Under this $SU(2)$ rotation the fermion field transforms
into
\begin{equation}
\label{pauligursey1}
\psi \rightarrow \psi' =
\hat{U}\psi\hat{U}^{\dag}=\frac{e^{i\varphi/2}}{\sqrt{1+|\tau|^2}}
[\psi-\tau^* \gamma_5 \psi^c]. 
\end{equation}
This is a  Pauli-G\"{u}rsey transformation with the parameters
\begin{eqnarray}
\label{pauligursey2}
a=\frac{e^{i\varphi/2}}{\sqrt{1+|\tau|^2}} \;\; ,\;\; b=\frac{-\tau^*
e^{i\varphi/2}}{\sqrt{1+|\tau|^2}}. 
\end{eqnarray}
Note that if the mass term is considered a pairing interaction the
Pauli-G\"{u}rsey transformation can be viewed as the associated
Bogoliubov transformation.  In light of Eqs. (\ref{pauligursey1}) and
(\ref{pauligursey2}) from now on we designate the $SU(2)$ algebra
spanned by $A_{\pm}$ and $A_0$ as $SU(2)_{\rm PG}$. It is also easy to
show that the $U(1)$ algebra spanned by $D_0$ commutes with
$SU(2)_{\rm PG}$.  

At this point the role of various components of the $Sp(4)$ algebra is
identified. It is easy to show that
\begin{equation}
D_0 = - \frac{1}{2}\int d^3{x}(\psi^{\dag} \gamma_5 \psi)
\end{equation}
Hence the $U(1)_{\chi}$ algebra spanned by $D_0$ is a chirality
transformation, the $SU(2)_{\rm PG}$ generates the Pauli-G\"{u}rsey
transformation and the most general neutrino mass Hamiltonian sits in
the $Sp(4)/SU(2)_{\rm PG} \times U(1)_{\chi}$ coset. It turns out that
the $SU(2)_{\rm PG}$ has another interesting function, namely that it
generates the see-saw mechanism. To illustrate this we first rewrite
Eq. (\ref{2.12}) in the form 
\begin{eqnarray}
\label{new2.12}
H_{m}=m_D(D_++D_-) &+& \left(\frac{m_L+m_R}{2}\right) \left[
(L_++L_-)+(R_++R_-) \right] \nonumber \\   &+& \left(
\frac{m_L-m_R}{2} \right) \left[ (L_++L_-)-(R_++R_-) \right].
\end{eqnarray}
One can easily show that under an $SU(2)_{\rm PG}$ rotation one gets 
\begin{equation}
\label{ltrans}
L_+ \rightarrow L_+' = \frac{e^{-i\varphi}}{1+|\tau|^2} \left[ L_+ +
\tau D_+ + \tau^2 R_+ \right];
\end{equation}
\begin{equation}
\label{rtrans}
R_+ \rightarrow R_+' = \frac{e^{i\varphi}}{1+|\tau|^2} \left[ R_+ -
\tau^* D_+ + (\tau^*)^2 L_+ \right];
\end{equation}
\begin{equation}
\label{dtrans}
D_+ \rightarrow D_+' = \frac{1}{1+|\tau|^2} \left[ (1-|\tau|^2) D_+  +
2 \tau R_+ - 2 \tau^* L_+ \right]. 
\end{equation}
From Eqs. (\ref{ltrans}) and (\ref{rtrans}) it follows that under an
$SU(2)_{\rm PG}$ rotation one has 
\begin{equation}
\label{ltotaltrans1}
\left[(L_++L_-)+(R_++R_-)\right] \rightarrow
\left[(L_+'+L_-')+(R_+'+R_-')\right] = \left[(L_++L_-)+(R_++R_-)\right]
\end{equation}
and
\begin{eqnarray}
\label{ltotaltrans2}
[ (L_++L_-) &-& (R_++R_-) ] \rightarrow
\left[(L_+'+L_-')-(R_+'+R_-')\right] \nonumber \\ &=&
\frac{1}{1+\tau^2} \left[ (1-\tau^2) \left( (L_++L_-)-(R_++R_-)\right)
+ 2 \tau \left( D_++D_- \right) \right] . 
\end{eqnarray}
In writing down Eqs. (\ref{ltotaltrans1}) and (\ref{ltotaltrans2}) we
took $\varphi=0$ and $\tau$ to be real assuming the masses in
Eq. (\ref{new2.12}) have no phases. If phases are included one should
restore $\varphi$ and a complex $\tau$. 

Under the $SU(2)_{\rm PG}$ rotation the mass Hamiltonian of
Eq. (\ref{new2.12}) transforms into
\begin{eqnarray}
\label{csaw}
H_m \rightarrow H_{m}' &=& \hat{U} H_{m} \hat{U}^{\dag} =
m_D(D_+'+D_-') + \left(\frac{m_L+m_R}{2}\right) \left[
(L_+'+L_-')+(R_+'+R_-') \right] \nonumber \\   &+& \left(
\frac{m_L-m_R}{2} \right) \left[ (L_+'+L_-')-(R_+'+R_-')
\right]\nonumber \\ &=& \left(\frac{m_L+m_R}{2}\right) \left[
(L_++L_-)+(R_++R_-) \right] \nonumber \\   &+& \frac{1}{1+\tau^2}
\left[ m_D (1 - \tau^2) + \left( m_L-m_R \right) \tau \right]
(D_++D_-)  \nonumber \\ &+& \frac{1}{1+\tau^2} \left[ - 2 m_D \tau +
\left( \frac{m_L-m_R}{2} \right) (1 - \tau^2) \right] \left[
(L_++L_-)-(R_++R_-)\right]
\end{eqnarray}
One can eliminate various terms from Eq. (\ref{csaw}) by a suitable
choice of $\tau$. For example, if one would like to eliminate the
Dirac mass term $(D_++D_-)$ one can choose $\tau$ to make its
coefficient zero. Introducing the angle $\delta$ 
\begin{equation}
\label{delta}
\tau = \tan \delta ,
\end{equation}
one finds that the choice 
\begin{equation}
\label{2delta}
\tan 2 \delta = - \frac{2 m_D}{(m_L-m_R)}, \> ,
\end{equation}
achieves this goal. The choice in Eq. (\ref{2delta}) corresponds to:
\begin{equation}
\label{3delta}
\cos 2 \delta = - \frac{(m_L-m_R)}{[4 m_D^2 + (m_L-m_R)^2 ]^{1/2}},
\>\>\> \sin  2 \delta =  \frac{2m_D}{[4 m_D^2 + (m_L-m_R)^2 ]^{1/2}},
\end{equation}
Note that this solution exists even in the limiting case of
$m_L=m_R$. 
Substituting Eq. (\ref{3delta}) into Eq. (\ref{csaw}) one obtains
\begin{eqnarray}
\label{1csaw}
H_m \rightarrow H_{m}' &=& \hat{U} H_{m} \hat{U}^{\dag} =
\left(\frac{m_L+m_R}{2}\right) \left[  (L_++L_-)+(R_++R_-) \right]
\nonumber \\ &-& \frac{1}{2} \left[  4 m_D^2 +  \left( m_L-m_R
\right)^2 \right]^{1/2}   \left[ (L_++L_-)-(R_++R_-)\right].
\end{eqnarray}
Making the choice $m_L=0$ and $m_R \gg m_D$ one obtains the standard
see-saw result
\begin{equation}
H_{m}' \approx m_R (R_++R_-) - \frac{m_D^2}{m_R} (L_++L_-).
\end{equation}

So far the discussion was for a single neutrino flavor. For three
neutrino flavors one can introduce three commuting copies of the
$Sp(4)$ algebra and the similar arguments follow. It is possible to
introduce a single $Sp(4)$ algebra for three flavors if the individual
masses of the mass eigenstates  are equal up to phases. This case
seems to be too restrictive for model building as it is at variance
with the recent observations. 
 
\section{Conclusions}

We showed that the neutrino mass problem can be formulated
algebraically. The algebra is Sp(4). A subgroup of this algebra
generates the see-saw mechanism and the neutrino mass Hamiltonian sits
in the leftover coset space. We should emphasize that even though in
writing down the symplectic algebra we utilized a framework which is
symmetric under the exchange of the left- and
right-handed fields, the algebraic basis we introduced does not
specify the values of $m_L$ and $m_R$. In particular the value $m_L=0$
is not excluded. Note that in the special case of 
$m_L = m_R$ and $m_D=0$, Eq. (\ref{ltotaltrans1}) indicates that a
general  Pauli-G\"{u}rsey transformation given in Eq. (\ref{3.3})
leaves the mass-term invariant. Indeed in this limit only the diagonal
$SU(2)_{L+R}$ subgroup of $SU(2)_L \times SU(2)_R$ (spanned by the
$L$'s and $R$'s respectively) is sufficient to describe the mass
Hamiltonian. This $SU(2)_{L+R}$ commutes with $SU(2)_{PG}$. 

Our analysis here concerns the neutrino mass, not the mixing matrix.
Recent observations of the atmospheric neutrinos \cite{atm} very
strongly suggest that muon neutrino maximally mixes with another
neutrino. In this case one can show that one of the mass eigenstates
decouples from the problem reducing the dimension of flavor space by
one \cite{bfmixing}, which suggests the existence of an underlying
symmetry.  It would be interesting to see if the $Sp(4)$ or a higher
symmetry could make a statement about the mixing matrix. Work along
this direction is currently in progress.

\section*{ACKNOWLEDGMENTS}

This work was supported in part by the U.S. National Science
Foundation Grant No.\ PHY-9605140 at the University of Wisconsin, and
in part by the University of Wisconsin Research Committee with funds
granted by the Wisconsin Alumni Research Foundation.  A.B.B.\
acknowledges the support of the Alexander von Humboldt-Stiftung.  The
work of N.\"{O} was in part supported by the Scientific and Technical
Research Council of Turkey (TUBITAK) under the project No. TBAG-1247. 
A.B.B. thanks to the Max-Planck-Institut f\"ur Kernphysik for the very
kind hospitality.


\begin{references}

\bibitem{atm}
  Y. Fukuda, {\em et al.}, (The Superkamiokande collaboration), Phys.
  Rev.  Lett.  {\bf 81}, 1562 (1998), {\bf 82}, 2644 (1999), 
  Phys. Lett. B {\bf 467}, 185 (1999). 

\bibitem{solar}
  B.T.  Cleveland, {\em et al.}, Astrophys. J.  {\bf 496}, 505 (1998);
  W. Hampel, {\em et al.} (GALLEX Collaboration), Phys. Lett. B {\bf
  420}, 114 (1998), {\bf 447}, 127 (1999);
  J.N. Abdurashitov, {\em et al.} (SAGE Collaboration),
  Phys. Rev. Lett. {\bf 83}, 4686 (1999), Phys. Rev. C {\bf 60},
  055801 (1999); Y. Fukuda, {\em et al.}, (The Superkamiokande
  collaboration), Phys. Rev. Lett. {\bf 82} 2430 (1999). 

\bibitem{seesaw}
  M. Gell-Mann, P. Ramond, and R.  Slansky, in {\em Supergravity\/},
  P. van Nieuwenhuizen and D.Z. Freedman, Eds., North Holland,
  Amsterdam, 1979, p.  315; T. Yanagida, in {\em Proceedings of the
  Workshop on the Unified Theory and Baryon Number in the Universe\/},
  O. Sawada and A. Sugamoto, Eds., (KEK Report 79-18, 1979), p.95; R.D. 
  Mohapatra and G. Senjanovic, Phys. Rev. Lett. {\bf 44}, 912 (1980). 

\bibitem{ramond}
  P. Ramond,  {\em Journeys Beyond the Standard Model}, Perseus Books,
  Cambridge (MA), 1999, p.227.

\bibitem{pere}
  A. Perelomov, {\em Generalized Coherent States and Their
  Applications}, Springer-Verlag, Berlin, 1986, p.111. 

\bibitem{boris}
  B. Kayser, {\em The Physics of Massive Neutrinos}, World Scientific,
  Singapore, 1989. 

\bibitem{haxste}
  W.C. Haxton and G.J. Stephenson, Prog. Part. Nucl. Phys. {\bf 12},
  409 (1984).      

\bibitem{murnaghan}
  F.D. Murnaghan, {\em The Theory of Group Representations}, Dover,
  New York, 1963. 

\bibitem{pauli}
  W. Pauli, Nuovo Cimento {\bf 6}, 204 (1957).

\bibitem{gursey}
  F. G\"{u}rsey, Nuovo Cimento {\bf 7}, 411 (1957). 

\bibitem{bfmixing}
  A.B. Balantekin and G.M. Fuller, Phys. Lett. B {\bf 471}, 195
  (1999). 

\end{references}
\end{document}